\documentclass{ws-ijmpa}

\begin{document}

\markboth{George Lazarides}
{Hybrid Inflation Followed by Modular Inflation}

\catchline{}{}{}{}{}

\title{HYBRID INFLATION FOLLOWED BY MODULAR
INFLATION}

\author{GEORGE LAZARIDES}

\address{Physics Division, School of Technology,
Aristotle University of Thessaloniki,\\
Thessaloniki 54124, Greece\\
lazaride@eng.auth.gr}

\maketitle

\begin{history}
\received{11 June 2007}
\end{history}

\begin{abstract}
Inflationary models with a superheavy scale
F-term hybrid inflation followed by an
intermediate scale modular inflation are
considered. The restrictions on the power
spectrum $P_{\cal R}$ of curvature perturbation
and the spectral index $n_{\rm s}$ from the
recent data within the power-law cosmological
model with cold dark matter and a cosmological
constant can be met provided that the number of
e-foldings $N_{\rm HI*}$ suffered by the pivot
scale $k_*=0.002/{\rm Mpc}$ during hybrid
inflation is suitably restricted. The
additional e-foldings needed for solving the
horizon and flatness problems are generated by
modular inflation with a string axion as
inflaton. For central values of
$P_{\cal R}$ and $n_{\rm s}$, the grand
unification scale comes out, in the case of
standard hybrid inflation, close to its
supersymmetric value $M_{\rm GUT}\simeq 2.86
\times 10^{16}~{\rm GeV}$, the relevant
coupling constant is relatively large
($\approx 0.005-0.14$), and
$10\lesssim N_{\rm HI*}\lesssim 21.7$. In the
shifted [smooth] hybrid inflation case, the
grand unification scale can be identified with
$M_{\rm GUT}$ for
$N_{\rm HI*}\simeq 21$ [$N_{\rm HI*}\simeq 18$].

\end{abstract}

\ccode{PACS number: 98.80.Cq}

\section{Introduction}

\par
Fitting the recent three-year results\cite{wmap3}
from the Wilkinson microwave anisotropy probe
satellite (WMAP3) with the standard power-law
cosmological model with cold dark matter and a
cosmological constant ($\Lambda$CDM), one
obtains\cite{wmap3} that, at the pivot scale
$k_*=0.002/{\rm Mpc}$,
\begin{equation}
\label{nswmap}
n_{\rm s}=0.958\pm0.016~\Rightarrow~0.926
\lesssim n_{\rm s}\lesssim 0.99
\end{equation}
at 95$\%$ confidence level. One of the most
natural and well-motivated classes of
inflationary models is the class\cite{hsusy} of
supersymmetric (SUSY) F-term hybrid inflation
(FHI)\cite{hybrid,susyhybrid} models. They
are realized at (or close to) the SUSY
grand unified theory (GUT) scale $M_{\rm GUT}
\simeq 2.86\times 10^{16}~{\rm GeV}$. However,
these inflationary models predict that the scalar
spectral index $n_{\rm s}$ is too close to unity
and without much running, which is in conflict
with the WMAP3 data. Moreover, including
supergravity (SUGRA) corrections with canonical
K\"{a}hler potential, $n_{\rm s}$
gets\cite{senoguz} closer to unity and can even
exceed it.

\par
One way out of this inconsistency
is\cite{lofti}\cdash\cite{rehman} to use a
quasi-canonical K\"{a}hler potential with a
convenient arrangement of the sign of one of its
terms. This yields\cite{king}\cdash\cite{gpp} a
negative mass term for the inflaton in the
inflationary potential, which, thus, in general
acquires a local maximum. Hilltop
inflation\cite{lofti} can then be realized as the
inflaton rolls from this maximum down to smaller
values. In this case, $n_{\rm s}$ can become
consistent with Eq.~(\ref{nswmap}), but only at
the cost of a mild tuning\cite{gpp} of the initial
conditions. Note, though, that, in some
cases,\cite{rehman,nsmooth} acceptable
$n_{\rm s}$'s can be obtained even without
this local maximum. Another possibility for
resolving the tension between FHI and the
data is\cite{battye} to include a small
contribution to the curvature perturbation from
cosmic strings,\cite{string,strings-domains}
which can make $n_{\rm s}$'s between 0.98 and 1
compatible with the data. However, the GUT scale
is constrained\cite{gpp,mairi,jp} to values well
below $M_{\rm GUT}$.

\par
In this talk, we present a recent
proposal\cite{mhin} of a two-step inflationary
set-up: a GUT scale FHI followed by an
intermediate scale modular inflation
(MI),\cite{modular} which allows acceptable
$n_{\rm s}$'s even with canonical K\"{a}hler
potential and without cosmic strings. The idea is
to constrain the number of e-foldings that $k_*$
suffers during FHI to relatively small values,
which reduces $n_{\rm s}$ to acceptable values.
The additional number of e-foldings required
for solving the horizon and flatness problems of
standard hot big bang cosmology is naturally
provided by MI, which can be easily realized by a
string axion. We show that this scheme can
satisfy all the relevant constraints with natural
values of its parameters.

\par
In Sec.~\ref{fhim}, we review the basic FHI
models. The calculation of their
inflationary observables is described in
Sec.~\ref{fhi}. Then, in Sec.~\ref{min}, we
sketch the main features of MI and, in
Sec.~\ref{cont}, we exhibit the constraints
imposed on our set-up. Finally, in
Sec.~\ref{num}, we present our numerical results
and, in Sec.~\ref{con}, we summarize our
conclusions.

\section{The FHI Models}
\label{fhim}

\par
The relevant superpotentials for the various
versions of FHI are\cite{hsusy}
\begin{equation}
\label{Whi}
W=\left\{
\begin{array}{rl}
\kappa S\left(\bar \Phi\Phi-M^2\right)\hfill
&~~~~~~~~~~\mbox{for standard FHI},\hfill\cr
\kappa S\left(\bar\Phi\Phi-M^2\right)-
S\frac{(\bar\Phi\Phi)^2}{M_{\rm S}^2}\hfill
&~~~~~~~~~~\mbox{for shifted FHI},\hfill\cr
S\left(\frac{(\bar\Phi\Phi)^2}{M_{\rm S}^2}-
\mu_{\rm S}^2\right)\hfill
&~~~~~~~~~~\mbox{for smooth FHI},\hfill\cr
\end{array}
\right.
\end{equation}
where $\bar{\Phi}$, $\Phi$ are left handed
superfields belonging to conjugate
representations of a GUT gauge group $G$ and
reducing its rank by their vacuum expectation
values (VEVs), $S$ is a gauge singlet left
handed superfield, $M_{\rm S}\sim 5\times10^{17}
~{\rm GeV}$ is the string scale, and $\kappa$ and
$M$, $\mu_{\rm S}~(\sim M_{\rm GUT})$ are made
real and positive by field redefinitions.

\par
The superpotential for
standard\cite{hybrid,susyhybrid} FHI in
Eq.~(\ref{Whi}) is the most general
renormalizable superpotential consistent with a
global ${\rm U}(1)$ R symmetry\cite{susyhybrid}
under which
\begin{equation}
\label{Rsym}
S\ \to\ e^{i\alpha}\,S,\quad\bar\Phi\Phi\ \to\
\bar\Phi\Phi.
\end{equation}
Note, in passing, that global continuous
symmetries such as this R symmetry can
effectively arise\cite{laz1} from the rich
discrete symmetry groups encountered in many
compactified string theories (see e.g.
Ref.~\refcite{laz2}). Including in the
superpotential for standard FHI the leading
non-renormalizable term, one obtains the
superpotential for shifted\cite{jean} FHI in
Eq.~(\ref{Whi}). The superpotential for
smooth\cite{pana1} FHI is produced by further
imposing a $Z_2$ symmetry under which
$\Phi\rightarrow -\Phi$ and, thus, allowing only
even powers of $\bar{\Phi}\Phi$.

\par
The vanishing of the D-terms implies that $\vert
\langle\bar{\Phi}\rangle\vert=\vert\langle\Phi
\rangle\vert$, while the vanishing of the F-terms
gives the VEVs of the fields in the SUSY vacuum,
namely $\langle S\rangle=0$ and $\vert\langle
\bar{\Phi}\rangle\vert=\vert\langle\Phi\rangle
\vert\equiv v_{_G}$ with
\begin{equation}
\label{vevs}
v_{_G}=\left\{
\begin{array}{rl}
M\hfill
&~~~~~~~~~~\mbox{for standard FHI},\hfill\cr
\frac{M}{\sqrt{2\xi}}\sqrt{1-\sqrt{1-4\xi}}\hfill
&~~~~~~~~~~\mbox{for shifted FHI},\hfill\cr
\sqrt{\mu_{\rm S}M_{\rm S}}\hfill
&~~~~~~~~~~\mbox{for smooth FHI}, \hfill
\cr
\end{array}
\right.
\end{equation}
where $\xi\equiv M^2/\kappa M_{\rm S}^2$ with
$1/7.2<\xi<1/4$.\cite{jean} So, the $W$'s in
Eq.~(\ref{Whi}) lead to the spontaneous breaking
of $G$. The same superpotentials give rise to
hybrid inflation. This is due to the fact that,
for large enough values of $|S|$, there exist
flat directions in field space,
i.e. valleys of local minima of the classical
potential with constant (or almost constant in
the case of smooth FHI) potential energy
density, which can serve as inflationary paths.

\par
The dominant contribution to the (inflationary)
potential energy density along these paths is
\begin{equation}
\label{V0}
V_{\rm HI0}=\left\{
\begin{array}{rl}
\kappa^2 M^4\hfill
&~~~~~~~~~~\mbox{for standard FHI},\hfill\cr
\kappa^2 M_\xi^4\hfill
&~~~~~~~~~~\mbox{for shifted FHI},\hfill\cr
\mu_{\rm S}^4\hfill
&~~~~~~~~~~\mbox{for smooth FHI},\hfill\cr
\end{array}
\right.
\end{equation}
where $M_\xi\equiv M\sqrt{1/4\xi-1}$. For
inflation to
be realized, we need a slope along the flat
direction (inflationary valley) to drive the
inflaton towards the vacuum. In the cases of
standard\cite{susyhybrid} and shifted\cite{jean}
FHI, this slope is generated by the SUSY
breaking on this valley caused by the
non-vanishing $V_{\rm HI0}$ on the valley. This
gives rise to logarithmic radiative corrections
to the potential. On the other hand, in the case
of smooth\cite{pana1} FHI, the inflationary
valley is not classically flat and, thus, there
is no need of radiative corrections. The relevant
correction $V_{\rm HIc}$ to the inflationary
potential can be written as follows:
\begin{equation}
\label{Vcor} V_{\rm HIc}=\left\{
\begin{array}{rl}
\frac{\kappa^4M^4{\sf N}}{32\pi^2}\left(2\ln\frac
{\kappa^2x M^2}{Q^2}+(x+1)^{2}\ln(1+x^{-1})
\!+\!(x-1)^{2}\ln(1-x^{-1})\right)\hfill\cr
\quad\quad\quad\quad\quad\quad\quad\quad\quad
\quad\quad\quad\quad\quad\quad\quad\quad\quad
\quad\quad\quad
\mbox{for standard FHI},\hfill\cr
\frac{\kappa^4 M_\xi^4}{16\pi^2}\left(2\ln\frac{2
\kappa^2x_\xi M_\xi^2}{Q^2}+(x_\xi+1)^{2}
\ln(1+x_\xi^{-1})\!+\!(x_\xi-1)^{2}
\ln(1-x_\xi^{-1})\right)\hfill\cr
\quad\quad\quad\quad\quad\quad\quad\quad\quad
\quad\quad\quad\quad\quad\quad\quad\quad\quad
\quad\quad\quad
\mbox{for shifted FHI},\hfill\cr
-2\mu_{\rm S}^6M_{\rm S}^2/27\sigma^4\hfill
\quad\quad\quad\quad\quad\quad\quad\quad\quad
\quad\quad\quad\quad\,\,\,
\mbox{for smooth FHI},\hfill\cr
\end{array}
\right.
\end{equation}
where $\sigma\equiv\sqrt{2}\vert S\vert$ is the
canonically normalized inflaton field, ${\sf N}$
is the dimensionality of the representations to
which $\bar{\Phi}$ and $\Phi$ belong in the case
of standard FHI, $Q$ is a renormalization scale,
$x\equiv |S|^2/M^2$, and
$x_\xi\equiv\sigma^2/M^2_\xi$.
For minimal K\"{a}hler potential, the leading
SUGRA correction to the inflationary potential
reads\cite{hybrid,senoguz,jp}
\begin{equation}
\label{Vsugra}
V_{\rm HIS}=V_{\rm HI0}\frac{\sigma^4}
{8m^4_{\rm P}},
\end{equation}
where $m_{\rm P}\simeq 2.44\times 10^{18}~
{\rm GeV}$ is the reduced Planck scale. In the
case of standard FHI, the
contribution\cite{sstad} to the inflationary
potential from the soft SUSY breaking terms is
negligibly small in our set-up due to the large
$\kappa$'s encountered (see Sec.~\ref{num}). This
contribution, in general, does not
have\cite{sstad} a significant effect in the
cases of shifted and smooth FHI too. All in all,
the general form of the potential which drives
the various versions of FHI reads
\begin{equation}
\label{Vol}
V_{\rm HI}\simeq V_{\rm HI0}+V_{\rm HIc}+
V_{\rm HIS}.
\end{equation}

\par
During standard FHI, both $\bar{\Phi}$ and $\Phi$
vanish and so the GUT gauge group $G$ is restored.
As a consequence, topological defects
such as cosmic
strings,\cite{string,strings-domains}
magnetic monopoles,\cite{monopole,monopoles} or
domain walls\cite{strings-domains,domain} will be
copiously produced\cite{pana1} via the Kibble
mechanism\cite{kibble} during the
spontaneous breaking of $G$ at the end of FHI if
they are predicted by this symmetry breaking.
This, which could lead to a cosmological
catastrophe in the cases of monopoles and walls,
is avoided in shifted and smooth FHI, since the
form of $W$ allows the existence of non-trivial
inflationary valleys along which $G$ is
spontaneously broken (with $\bar\Phi$ and $\Phi$
acquiring non-zero values). Therefore, no
topological defects are produced in these cases.
In standard FHI, on the other hand, we must
normally ensure that no monopoles or walls are
predicted by the underlying particle physics
scheme. In our set-up, however, this restriction
can be evaded since the subsequent MI dilutes the
topological defects.

\section{The Dynamics of FHI}
\label{fhi}

\par
We will assume that all the cosmological scales
cross outside the horizon during FHI and do not
re-enter the horizon before the onset of MI (see
below). The latter guarantees
that they are not ``re-processed'' by MI. So, we
can apply the standard formalism (see e.g.
Ref.~\refcite{lectures}) to calculate the
inflationary observables of FHI. Namely, the
number of e-foldings $N_{\rm HI*}$ that the pivot
scale $k_*$ suffers during FHI
is given by
\begin{equation}
\label{Nefold}
N_{\rm HI*}=\:\frac{1}{m^2_{\rm P}}\;
\int_{\sigma_{\rm f}}^{\sigma_{*}}\,
d\sigma\: \frac{V_{\rm HI}}{V'_{\rm HI}},
\end{equation}
where the prime denotes derivation with respect
to $\sigma$, $\sigma_{*}$ is the value
of $\sigma$ when $k_*$ crosses outside the
horizon of FHI, and $\sigma_{\rm f}$ is the value
of $\sigma$ at the end of FHI. In the slow-roll
approximation, $\sigma_{\rm f}$ is found from the
condition
\begin{eqnarray}
\label{slow} &&\quad{\sf max}\{\epsilon(
\sigma_{\rm f}),|\eta(\sigma_{\rm f})|\}=1,\quad
\mbox{where}\quad\nonumber\\
&&\epsilon\simeq\frac{m^2_{\rm P}}{2}\left(
\frac{V'_{\rm HI}}{V_{\rm HI}}\right)^2\quad
\mbox{and}\quad\eta\simeq m^2_{\rm P}~\frac{
V''_{\rm HI}}{V_{\rm HI}}.
\end{eqnarray}
In standard\cite{susyhybrid} and
shifted\cite{jean} FHI, the end of
inflation coincides with the onset of the GUT
phase transition, i.e. the slow-roll conditions
are violated infinitesimally close to the
critical point at $\sigma=\sigma_{\rm c}\equiv
\sqrt{2}M$ [$\sigma=\sigma_{\rm c}\equiv M_\xi$]
for standard [shifted] FHI, where the
inflationary path is destabilized and the
``waterfall'' regime commences. On the contrary,
the end of smooth\cite{pana1} FHI is not
abrupt since the inflationary path is stable with
respect to variations in $\bar\Phi$, $\Phi$ for
all $\sigma$'s and $\sigma_{\rm f}$ is
found from Eq.~(\ref{slow}).

\par
The power spectrum $P_{\cal R}$ of the curvature
perturbation at $k_{*}$ is given by
\begin{equation}
\label{Pr}
P^{1/2}_{\cal R}=\: \frac{1}{2\sqrt{3}\,
\pi m^3_{\rm P}}\;
\left.\frac{V_{\rm HI}^{3/2}}{|V'_{\rm
HI}|}\right\vert_{\sigma=\sigma_*}.
\end{equation}
Finally, the spectral index $n_{\rm s}$ and its
running $dn_{\rm s}/d\ln k$ are
\begin{eqnarray}
\label{nS}
n_{\rm s}&=&1-6\epsilon(\sigma_*)\ +\
2\eta(\sigma_*)\quad\mbox{and}
\nonumber\\
dn_{\rm s}/d\ln k&=&2\left(4\eta(\sigma_*)^2
-(n_{\rm s}-1\right)^2)/3-2\xi(\sigma_*)
\end{eqnarray}
respectively with $\xi\simeq m_{\rm P}^4~
V'_{\rm HI} V'''_{\rm HI}/V^2_{\rm HI}$.

\section{The Basics of MI}
\label{min}

After the gravity mediated soft SUSY breaking,
the potential for MI is\cite{modular}
\begin{equation}
V_{\rm MI}=V_{\rm MI0}-\frac{1}{2}m_s^2s^2+\dots,
\label{Vinf}
\end{equation}
where $s$ is the canonically normalized real
string axion field, the ellipsis denotes terms
which stabilize $V_{\rm MI}$ at
\mbox{$s\sim m_{\rm P}$},
\begin{equation}
V_{\rm MI0}=v_s(m_{3/2}m_{\rm P})^2,\quad
{\rm and}\quad m_s\sim m_{3/2}
\label{Vm}
\end{equation}
with $m_{3/2}\sim 1~{\rm TeV}$ being the
gravitino mass and the dimensionless parameter
$v_s$ being of order unity, which yields
$V_{\rm MI0}^{1/4}\simeq 3\times 10^{10}~
{\rm GeV}$. In this model, inflation can be of
the fast-roll type.\cite{fastroll} The field
evolution is given\cite{fastroll} by
\begin{equation}
s\simeq s_{\rm i}e^{F_s \Delta N_{\rm MI}}\quad
{\rm with}\quad
F_s\equiv\sqrt{\frac{9}{4}+\left(\frac{m_s}
{H_s}\right)^2}-\frac{3}{2},
\label{Fs}
\end{equation}
where $s_{\rm i}$ is the initial value of $s$ (at
the onset of MI), $H_s\simeq\sqrt{V_{\rm MI0}}/
\sqrt{3}m_{\rm P}$ is the Hubble parameter
corresponding to $V_{\rm MI0}$, and
$\Delta N_{\rm MI}$ is the number of e-foldings
obtained from $s=s_{\rm i}$ until a given $s$.

\par
From Eq.~(\ref{Fs}), we estimate the number
of e-foldings $N_{\rm MI}$ during MI:
\begin{equation}
N_{\rm MI}\simeq\frac{1}{F_s}\ln\left(
\frac{s_{\rm f}}{s_{\rm i}}\right),
\label{Nmp}
\end{equation}
where $s_{\rm f}={\sf min}\{\langle s\rangle,
s_{\rm sr}\}$ is the final value of $s$ with
$\langle s\rangle\sim m_{\rm P}$ being the VEV
of $s$ and $s_{\rm sr}$ determined by the
condition
\begin{equation}
\label{varepsilon}
\epsilon_{\rm MI}\equiv -\frac{\dot{H}_{\rm MI}}
{H^2_{\rm MI}}\simeq\frac{1}{2}F_s^2\left(
\frac{s}{m_{\rm P}}\right)^2=1
\end{equation}
($H_{\rm MI}$ is the Hubble parameter during MI
and the dot denotes derivation with respect to
the cosmic time). For definiteness, we take
$\langle s\rangle=m_{\rm P}$ throughout our
calculation.

\section{Observational Constraints}
\label{cont}

Our scenario needs to satisfy the following
constraints:

\begin{romanlist}[(ii)]

\item
The power spectrum in Eq.~(\ref{Pr}) is to be
confronted with the WMAP3 data\cite{wmap3}:
\begin{equation}
\label{Prob}
P^{1/2}_{\cal R}\simeq\: 4.86\times
10^{-5}\quad\mbox{at}\quad k_*=0.002/{\rm Mpc}.
\end{equation}

\item
In our case, the horizon and flatness problems
of big bang cosmology can be resolved provided
that the total number of e-foldings
$N_{\rm tot}$ suffered by $k_*$ is
given\cite{hybrid,anupam} by
\begin{equation}
N_{\rm tot}\simeq22.6+\frac{2}{3}
\ln\frac{V^{1/4}_{\rm HI0}}{{1~{\rm GeV}}}+
\frac{1}{3}\ln\frac{T_{\rm Mrh}}
{1~{\rm GeV}},
\label{Ntott}
\end{equation}
where $T_{\rm Mrh}$ is the reheat temperature
after the completion of MI. Here, we have assumed
that the reheat temperature after FHI is lower
than $V_{\rm MI0}^{1/4}$ and, thus, the whole
inter-inflationary period is matter dominated.
In our set-up, $N_{\rm tot}$ consists of two
contributions:
\begin{equation}
N_{\rm tot} =\: N_{\rm HI*}+N_{\rm MI}\,.
\label{Ntot}
\end{equation}

\item
The assumption that all the cosmological scales
leave the horizon during FHI and do not re-enter
the horizon before the onset of MI
yields\cite{anupam,astro} the restriction:
\begin{equation}
N_{\rm HI*}\gtrsim N^{\rm min}_{\rm HI*}
\simeq 3.9+\frac{1}{6}\ln\frac{V_{\rm HI0}}
{V_{\rm MI0}}.
\label{ten}
\end{equation}
The first term in the expression for
$N^{\rm min}_{\rm HI*}$ is the number of
e-foldings elapsed between the horizon crossing
of the pivot scale $k_*$ and the scale
$0.1/{\rm Mpc}$ during FHI. Length scales
$\sim10~{\rm Mpc}$ are starting to feel
non-linear effects and it is, thus, difficult to
constrain\cite{astro} primordial density
fluctuations on smaller length scales. So, we
take the largest cosmological scale to be about
$0.1/{\rm Mpc}$.

\item
In the FHI models, $|dn_{\rm s}/d\ln k|$
increases\cite{espinoza} as $N_{\rm HI*}$
decreases. Therefore, consistency with the
assumptions of the power-law $\Lambda$CDM
cosmological model, which requires that
\begin{equation}
\vert dn_{\rm s}/d\ln k\vert \ll 0.01,
\label{dnsk}
\end{equation}
yields a lower bound on $N_{\rm HI*}$. In our
numerical investigation (see Sec.~\ref{num}), we
display boundary curves for $dn_{\rm s}/d\ln k=
-0.005$ and $-0.01$.

\item
The requirement of naturalness of MI constrains
the dimensionless parameter $v_s$ in
Eq.~(\ref{Vm}) as follows:
\begin{equation}
\label{nin}
0.5\leq v_s\leq 10\quad\Rightarrow\quad 2.45
\gtrsim m_s/H_s\gtrsim 0.55,
\end{equation}
where we take $m_s=m_{3/2}$ (see below). The
lower bound on $v_s$ guarantees that the sum of
the two explicitly displayed terms in the right
hand side of Eq.~(\ref{Vinf}) is positive
for $s<m_{\rm P}$. From Eq.~(\ref{varepsilon}),
we see that, for the values of $m_s/H_s$ in
Eq.~(\ref{nin}), $s_{\rm sr}>m_{\rm P}$ and,
thus, $s_{\rm f}=m_{\rm P}$.
Eqs.~(\ref{Fs})--(\ref{varepsilon}) are not very
accurate near the upper bound on $m_s/H_s$ since,
in this region, the value of $\epsilon_{\rm MI}$
at $s=m_{\rm P}$ gets too close to unity and,
thus, the Hubble parameter does not remain
constant as $s$ approaches $m_{\rm P}$. So, our
results at large values of $m_s/H_s$ should be
considered only as indicative. Fortunately, the
interesting solutions with $n_{\rm s}$ near its
central value and, in the smooth FHI case,
$v_{_G}\sim M_{\rm GUT}$ lie near the lower bound
on $m_s/H_s$, where the accuracy of these
formulas is much better. Moreover,
\begin{equation}
\eta_{\rm MI}\equiv m_{\rm P}^2
\frac{V_{\rm MI}^{(2)}}{V_{\rm MI}}
\simeq -\frac{1}{3}\left(\frac{m_s}{H_s}
\right)^2\lesssim 1\quad\mbox{for}\quad
m_s/H_s\lesssim 1.73,
\end{equation}
where we again take $m_s=m_{3/2}$ and the
superscript $(n)$ denotes the $n$-th derivative
with respect to $s$. So, the interesting
solutions
correspond to slow- rather than fast-roll MI. The
unspecified terms in the ellipsis in the right
hand side of Eq.~(\ref{Vinf}) also generate an
uncertainty in
Eqs.~(\ref{Fs})--(\ref{varepsilon}), which will
be assumed negligible.

\item
Finally, we assume that FHI lasts long enough
so that the almost massless string axion field
$s$ is completely randomized\cite{randomize} by
its quantum fluctuations from FHI. We further
assume that
\begin{equation}
\label{random}
V_{\rm MI0}\lesssim H^4_{\rm HI0},
\end{equation}
where $H_{\rm HI0}=\sqrt{V_{\rm HI0}}/\sqrt{3}
m_{\rm P}$ is the Hubble parameter corresponding
to $V_{\rm HI0}$, so that all the values of $s$
belong to the randomization
region.\cite{randomize} The field
$s$ remains practically frozen during the
inter-inflationary period since the Hubble
parameter is larger than its mass. So, all the
initial values $s_{\rm i}$ of $s$ from zero to
$m_{\rm P}$ are equally probable. However, we
take $s_{\rm i}\gg H_{\rm HI0}/2\pi$ so that the
homogeneity of our present universe is not
jeopardized by the quantum fluctuations of $s$
from FHI. Randomization of the value of a scalar
field via inflationary quantum fluctuations
requires that this field remains almost massless
during inflation. For this, it is important that
the field does not acquire\cite{hybrid,effmass}
mass of the order of the Hubble parameter via the
SUGRA scalar potential. This is, indeed, the case
for the string axion during FHI (and the
inter-inflationary period).

\end{romanlist}

\begin{figure}[t!]
\centerline{
\psfig{file=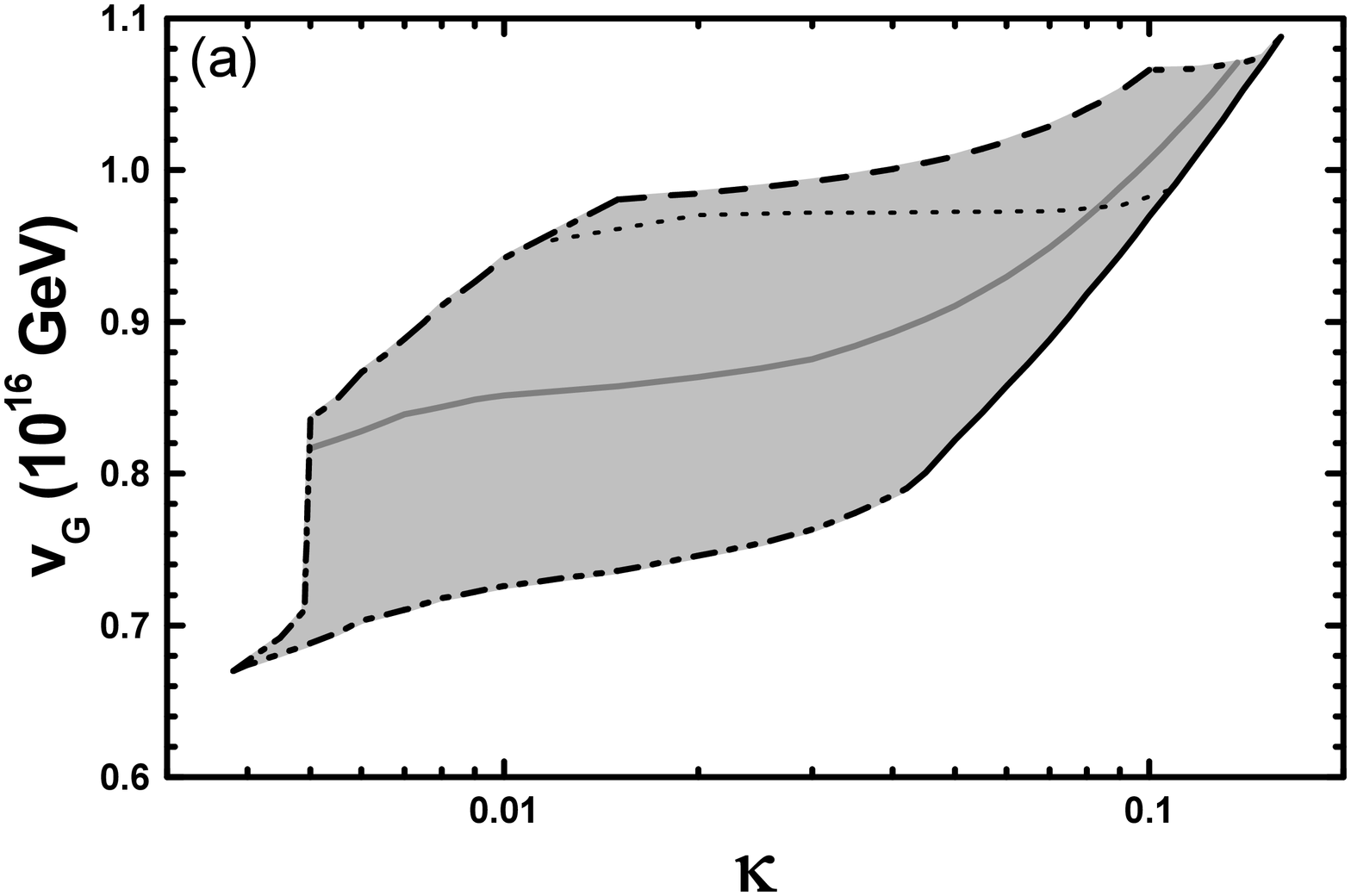,width=4.4cm,angle=-90}
\psfig{file=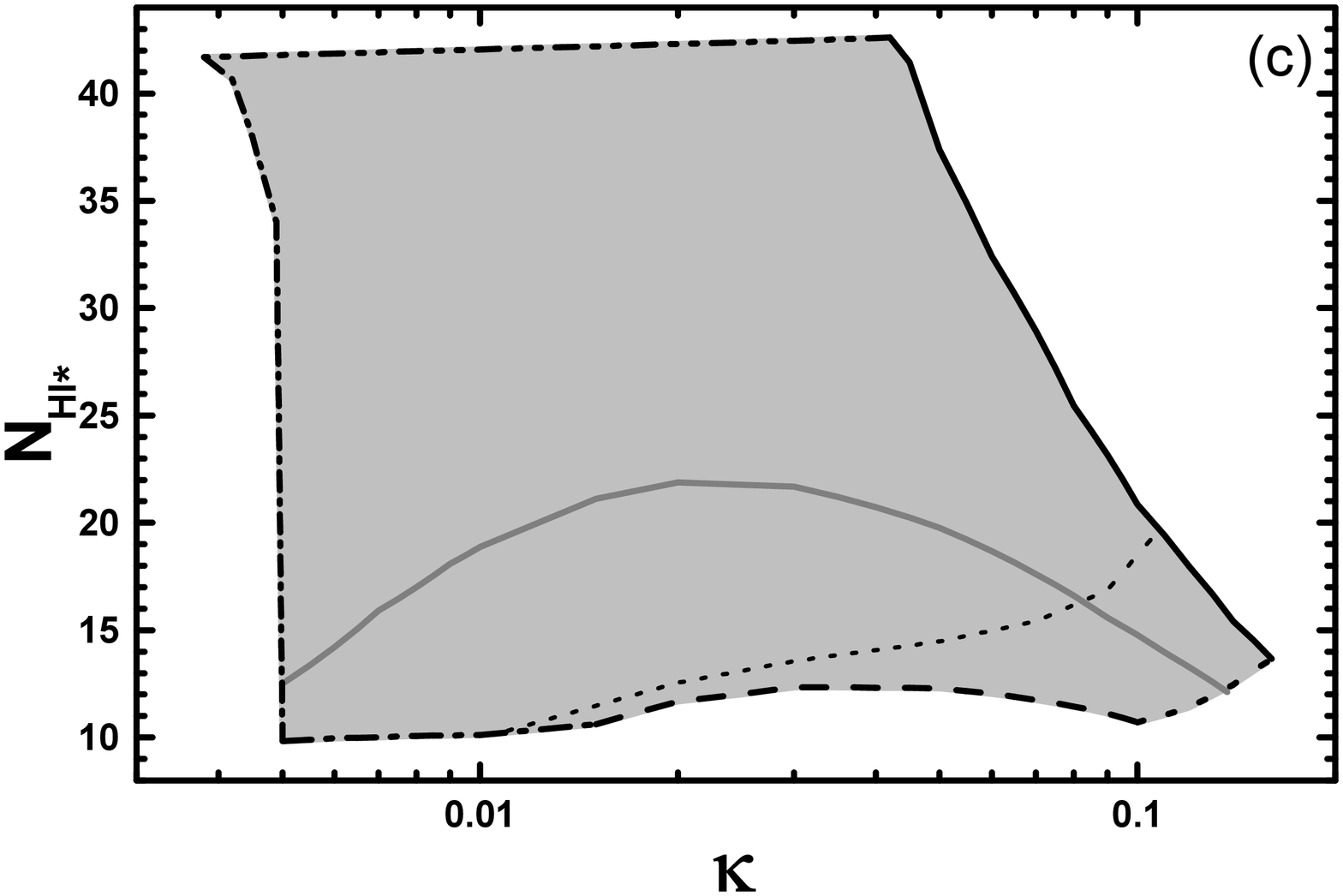,width=4.4cm,angle=-90}}
\centerline{
\psfig{file=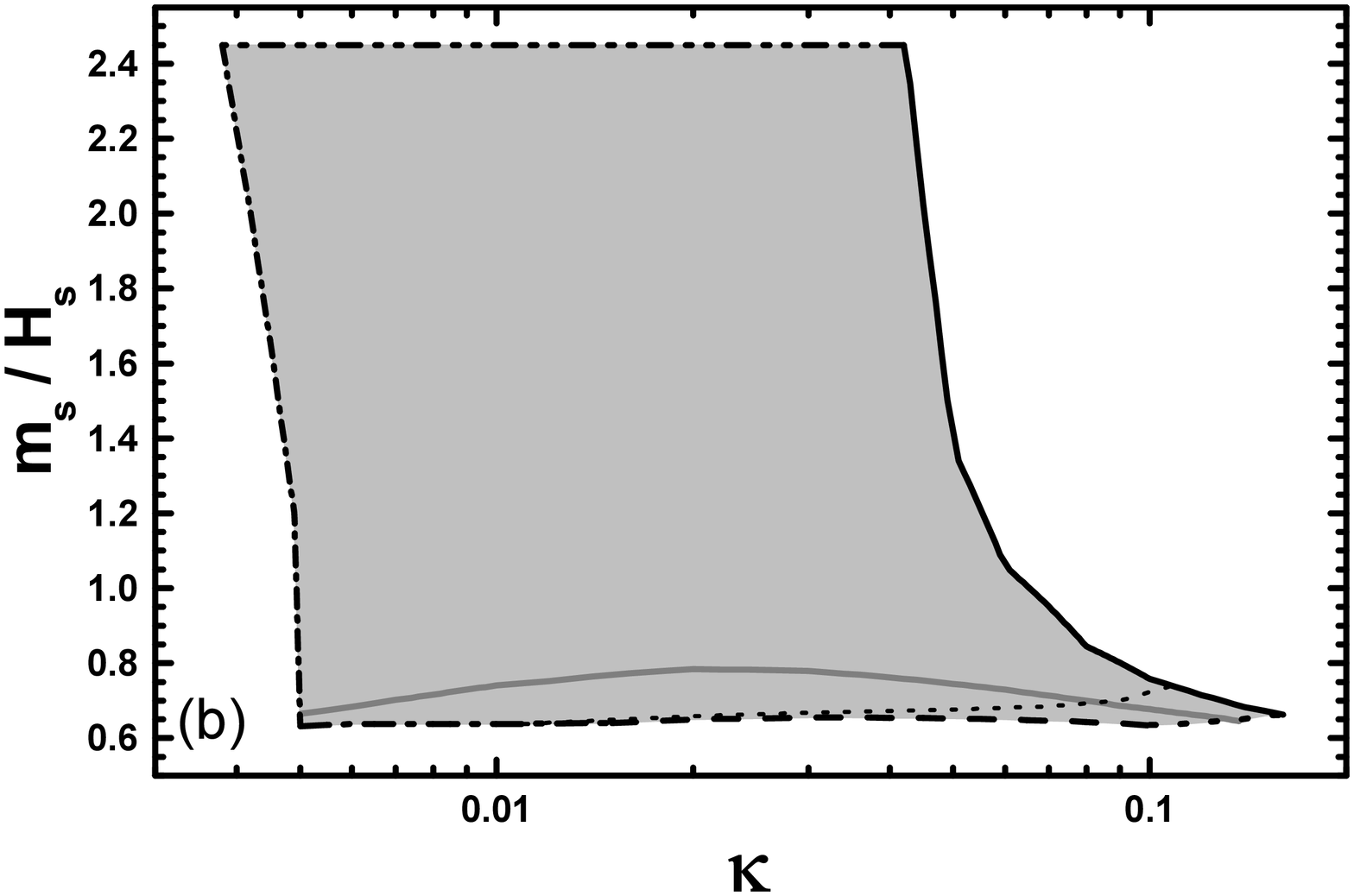,width=4.4cm,angle=-90}
\psfig{file=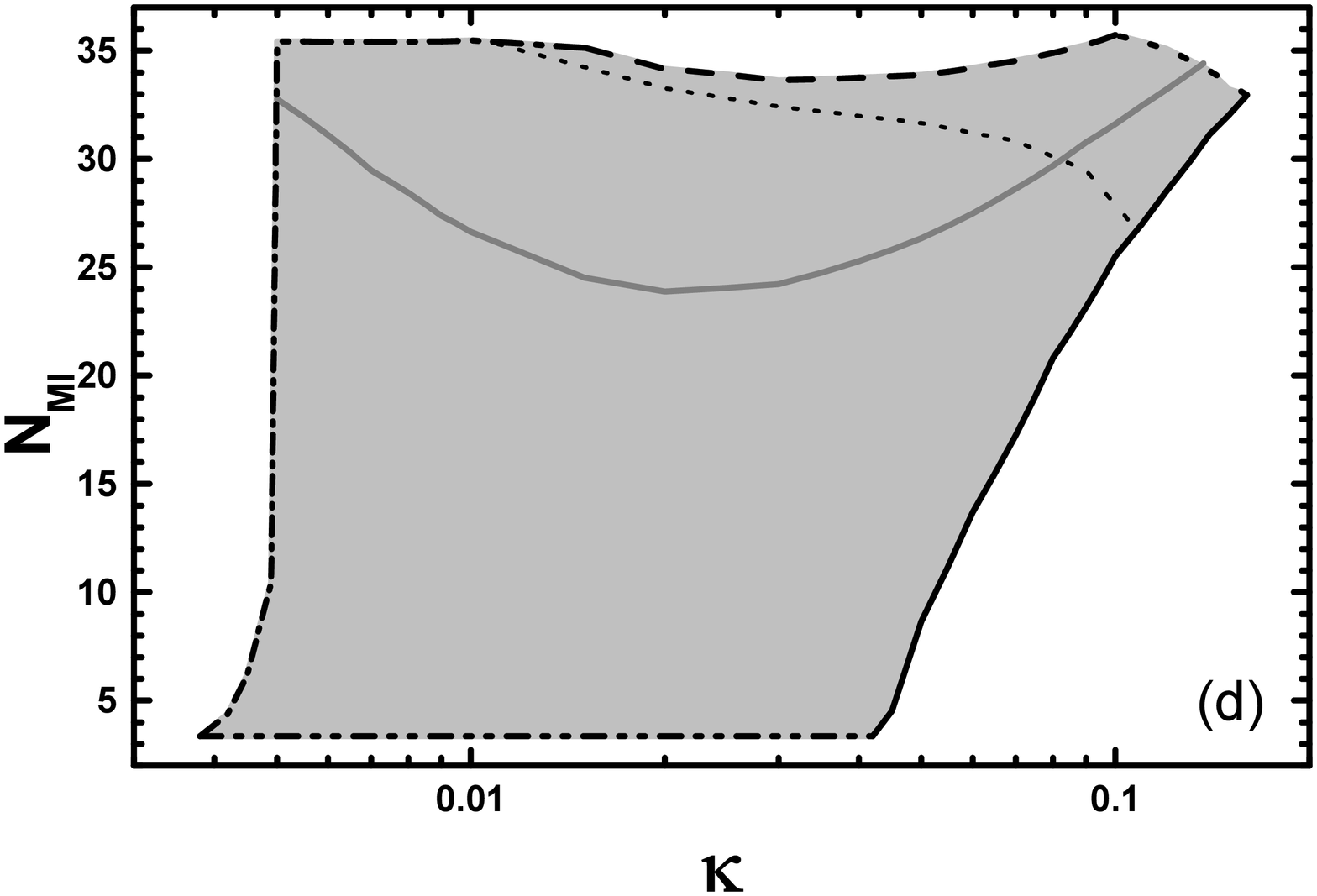,width=4.4cm,angle=-90}}
\vspace*{8pt}
\caption{Allowed (lightly gray shaded) regions in
the {(a)} $\kappa-v_{_{G}}$,
{(b)} $\kappa-m_s/H_s$, {(c)} $\kappa-N_{\rm HI*}$,
and {(d)} $\kappa-N_{\rm MI}$ plane for standard
FHI. The black solid [dashed] lines correspond to
the upper [lower] bound on $n_{\rm s}$ in
Eq.~(\ref{nswmap}), whereas the gray solid lines
correspond to its central value in this equation.
The dot-dashed [double dot-dashed] lines
correspond to the lower [upper] bound on
$N_{\rm HI*}~[m_s/H_s]$ from Eq.~(\ref{ten})
[Eq. (\ref{nin})]. The bold [faint] dotted lines
correspond to $dn_{\rm s}/d\ln k=-0.01$
[$dn_{\rm s}/d\ln k=-0.005$]. Finally, the lower
bound on $V_{\rm HI0}$ from Eq.~(\ref{random}) is
represented by the short dash-dotted lines.
\label{stad}}
\end{figure}

\section{Numerical Results}
\label{num}

For standard\cite{susyhybrid} FHI, we take
${\sf N}=2$, which
corresponds to the left-right symmetric GUT
gauge group ${\rm SU(3)_c\times SU(2)_L\times
SU(2)_R\times U(1)_{B-L}}$ with $\bar\Phi$ and
$\Phi$ being ${\rm SU(2)_R}$ doublets with
$B-L=-1$ and $1$ respectively. No cosmic strings
are produced\cite{trotta} during this realization
of standard FHI, which liberates the model from
extra restrictions on its parameters (for such
restrictions, see e.g. Refs.~\refcite{mairi,jp}).
For shifted\cite{jean} FHI, the GUT gauge group
is the Pati-Salam group\cite{ps}
${\rm SU(4)_c\times SU(2)_L\times SU(2)_R}$. This
predicts the existence of doubly
charged\cite{magg} magnetic monopoles which are,
though, not produced at the end of inflation as
mentioned in Sec.~\ref{fhim}. We
take $T_{\rm Mrh}=1~{\rm GeV}$ and $m_{3/2}=m_s=
1~{\rm TeV}$ throughout. These are indicative
values, which do not affect crucially our results.
Finally, we choose the initial value $s_{\rm i}$
of the string axion $s$ at the onset of MI to be
given by $s_{\rm i}=0.01\,m_{\rm P}$ in all the
cases that we consider. This value is close
enough to $m_{\rm P}$ to have a non-negligible
probability to be achieved by the randomization
of $s$ during FHI. At the same time, it is
adequately smaller than $m_{\rm P}$ to guarantee
good accuracy of
Eqs.~(\ref{Fs})--(\ref{varepsilon}) near the
interesting solutions and justify the fact that
we neglect the uncertainty from the ellipsis in
Eq.~(\ref{Vinf}). Moreover, larger $s_{\rm i}$'s
lead to smaller parameter space for interesting
solutions (with $n_{\rm s}$ near its central
value).

\par
Our input parameters are $\kappa$ (for standard
and shifted FHI with fixed $M_{\rm S}=5\times
10^{17}~{\rm GeV}$) or $M_{\rm S}$ (for smooth
FHI) and $\sigma_*$. Using Eqs.~(\ref{nS}) and
(\ref{Prob}), we extract $n_{\rm s}$ and $v_{_G}$
respectively. For every chosen $\kappa$ or
$M_{\rm S}$, we then restrict $\sigma_*$ so as to
achieve $n_{\rm s}$ in the range of
Eq.~(\ref{nswmap}) and take the output values
of $N_{\rm HI*}$. Finally, we find, from
Eqs.~(\ref{Ntott}) and (\ref{Ntot}), the required
$N_{\rm MI}$ and the corresponding $v_s$ or
$m_s/H_s$ from Eq.~(\ref{Nmp}).

\par
Our numerical results for the three versions of
FHI are presented in
Figs.~\ref{stad}--\ref{smth}.
In Fig.~\ref{shth}(a) [Fig.~\ref{smth}(a)], we
focus on a limited range of $\kappa$'s
[$M_{\rm S}$'s] for the sake of clarity of
the presentation. Let us discuss each case
separately:

\begin{figure}[t!]
\centerline{
\psfig{file=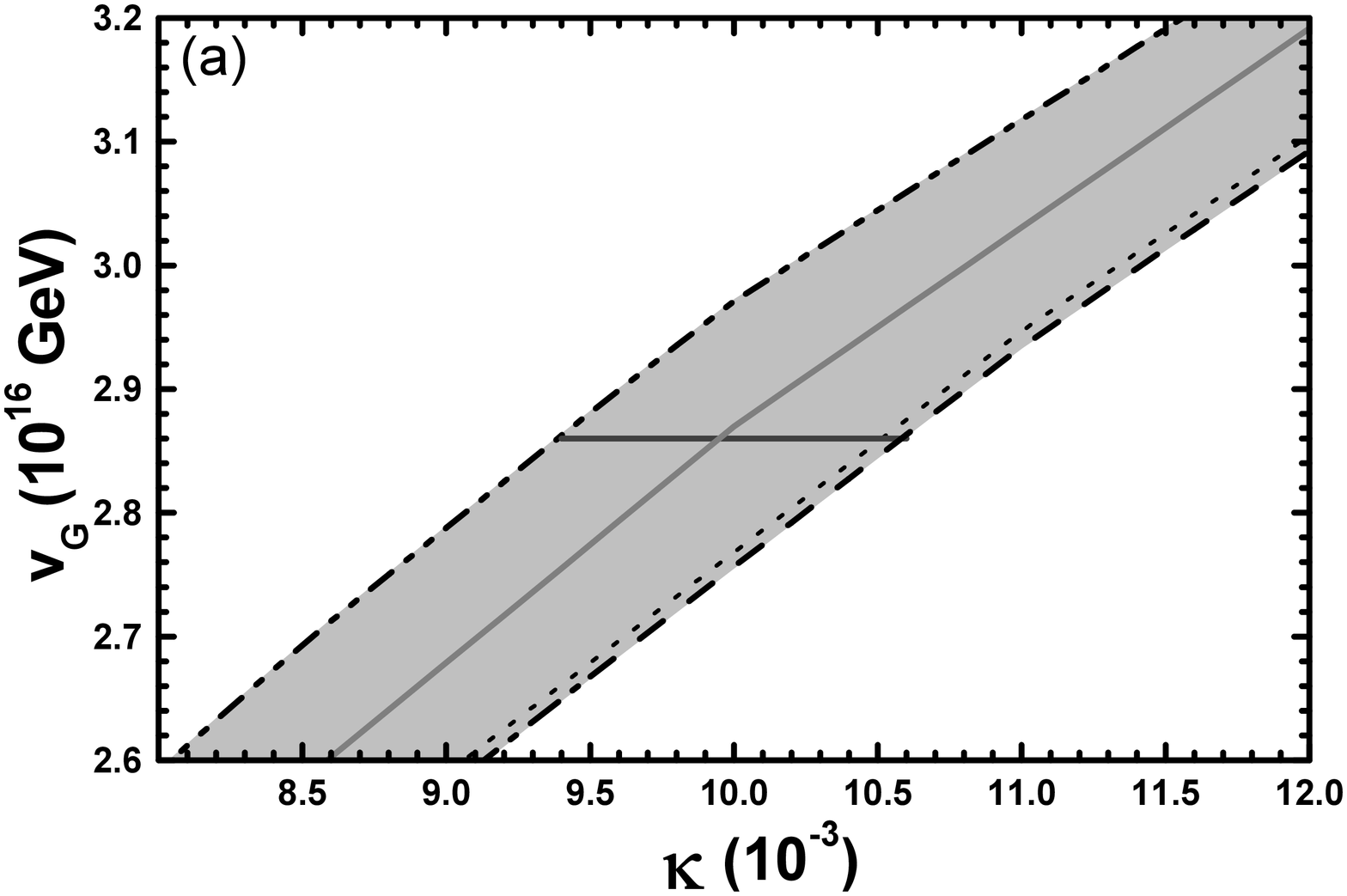,width=4.4cm,angle=-90}
\psfig{file=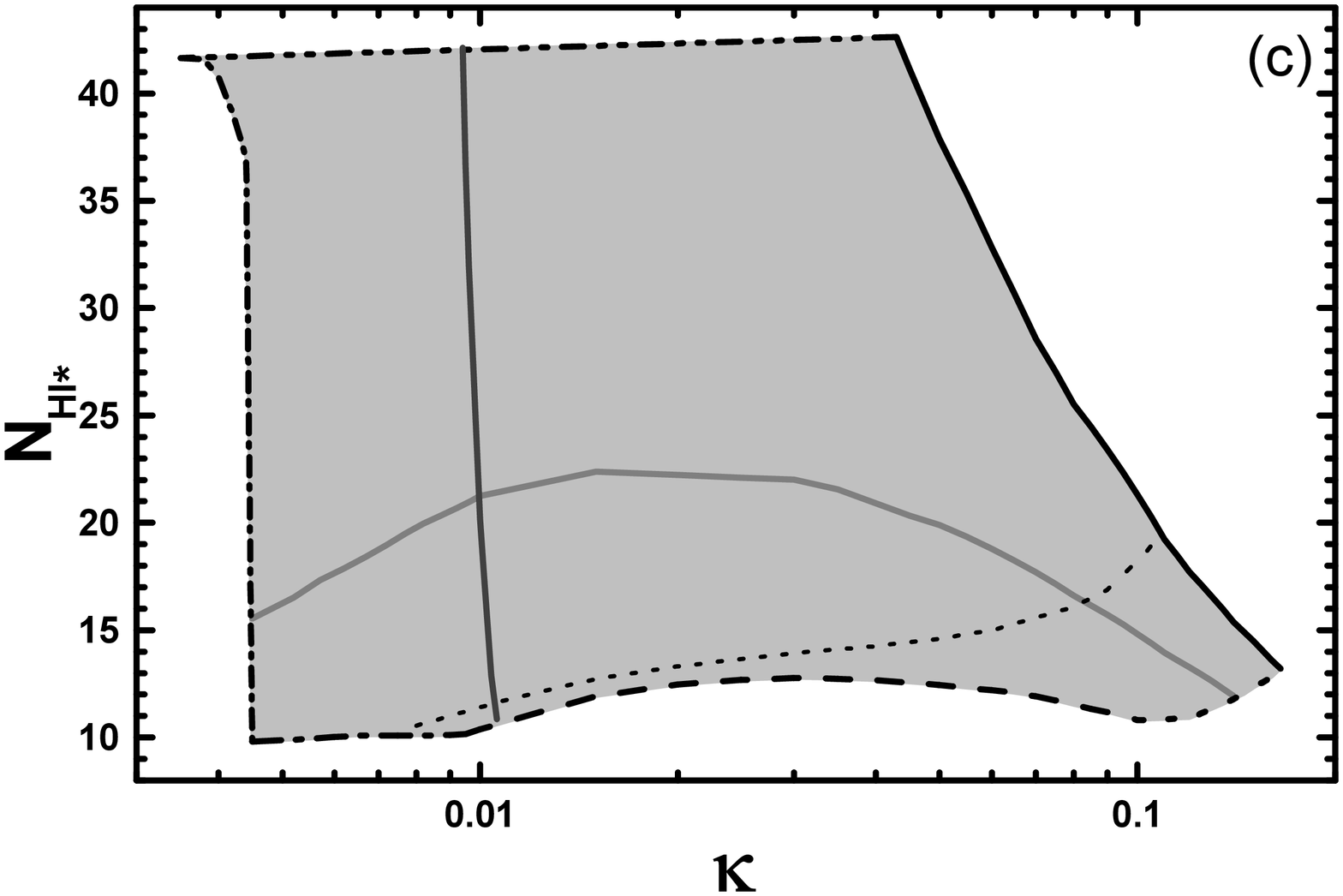,width=4.4cm,angle=-90}}
\centerline{
\psfig{file=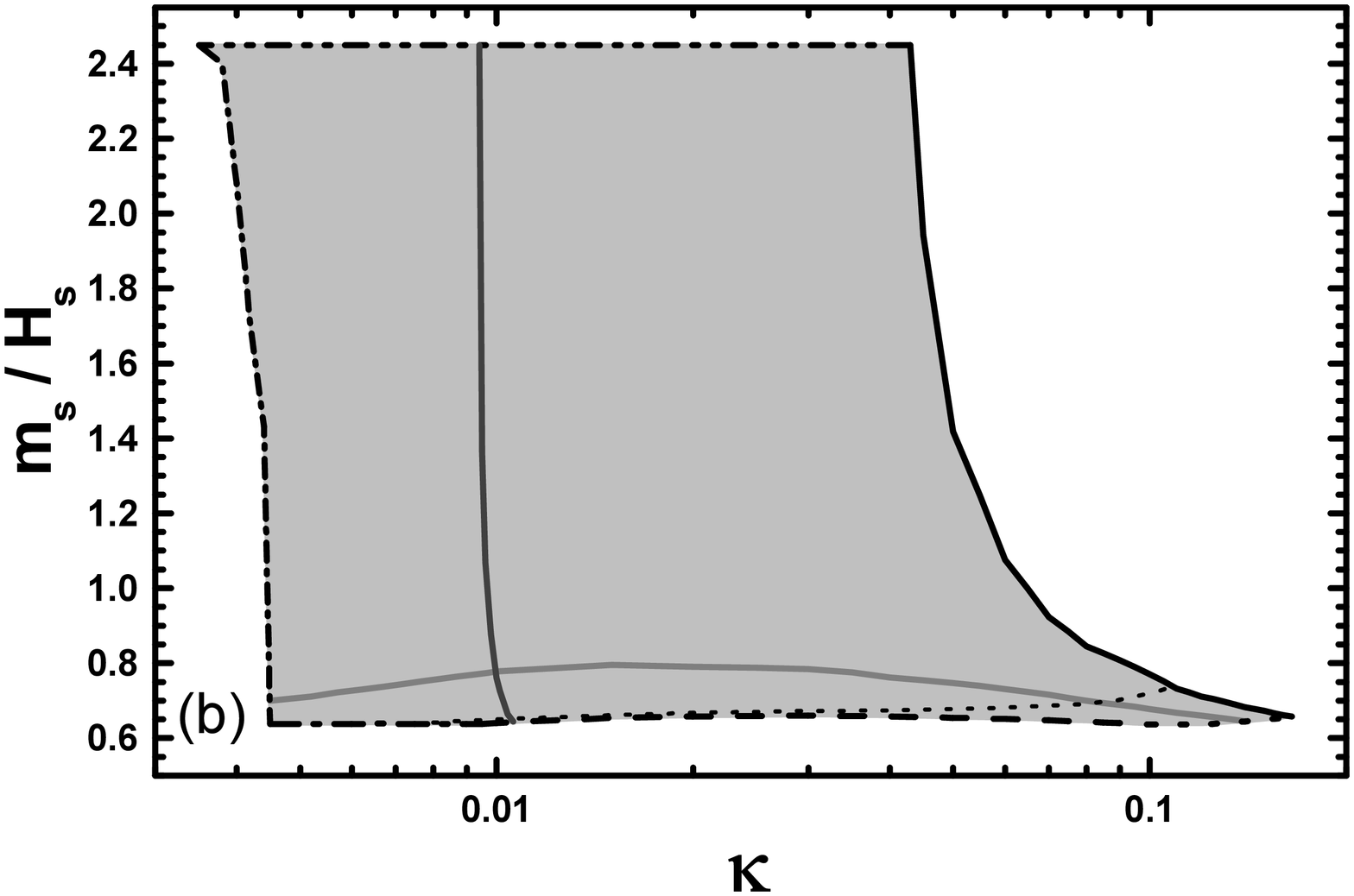,width=4.4cm,angle=-90}
\psfig{file=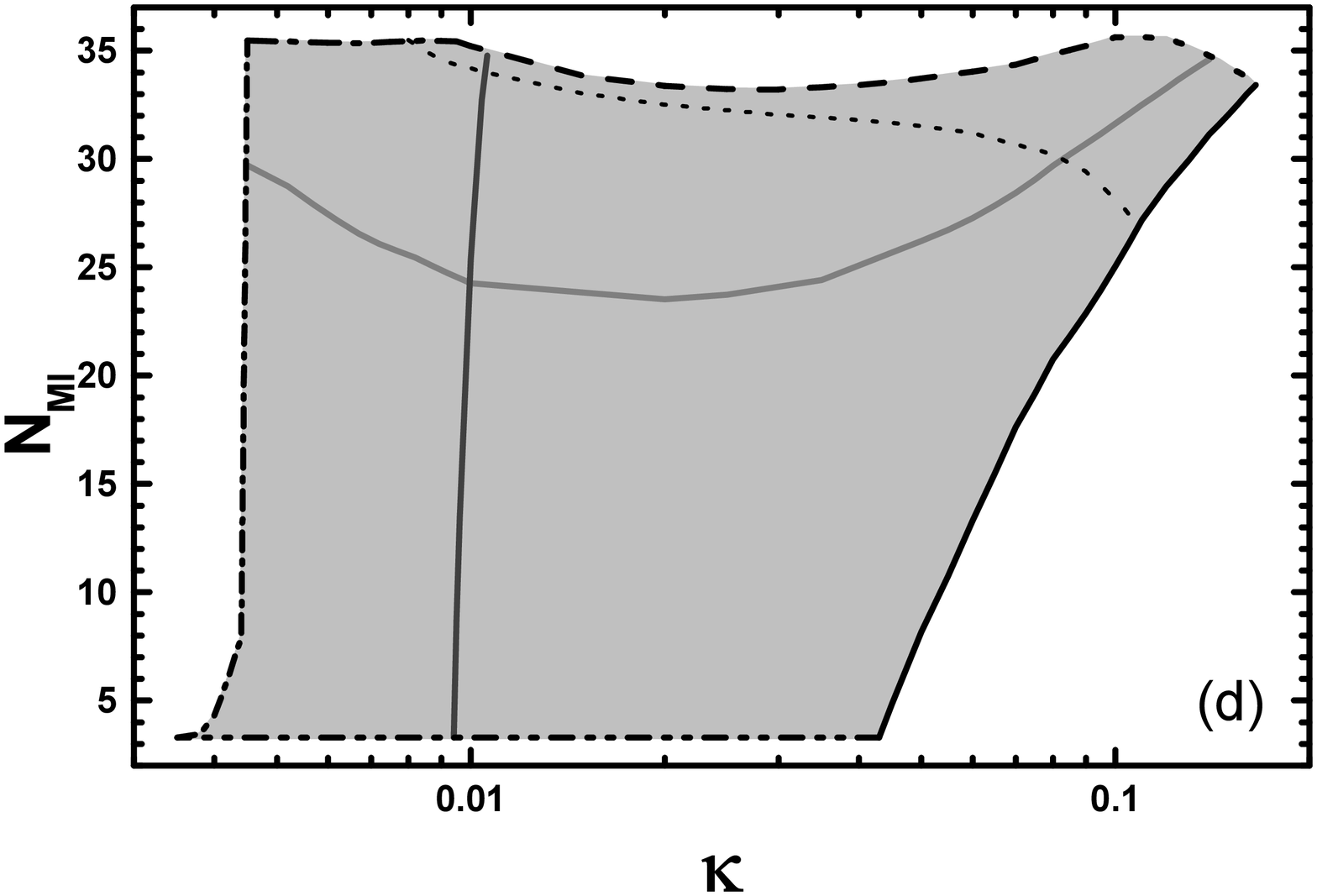,width=4.4cm,angle=-90}}
\vspace*{8pt}
\caption{Allowed regions in the
{(a)} $\kappa-v_{_G}$, {(b)} $\kappa-m_s/H_s$,
{(c)} $\kappa-N_{\rm HI*}$, and
{(d)} $\kappa-N_{\rm MI}$ plane for shifted FHI
with $M_{\rm S}=5\times10^{17}~{\rm GeV}$. Same
notation as in Fig.~\ref{stad}. We also include
dark gray solid lines corresponding to
$v_{_{G}}=M_{\rm GUT}$.}
\label{shth}
\end{figure}

\subsection{Standard FHI}

In Fig.~\ref{stad}, we present the regions
allowed by Eqs.~(\ref{nswmap}),
(\ref{Prob})--(\ref{nin}), and (\ref{random}) in
the {(a)} $\kappa-v_{_G}$, {(b)} $\kappa-m_s/H_s$,
{(c)} $\kappa-N_{\rm HI*}$, and {(d)}
$\kappa-N_{\rm MI}$ plane for standard FHI. We
observe the following:

\begin{romanlist}[(ii)]

\item
The resulting $v_{_G}$'s and $\kappa$'s are
restricted to rather large values compared to
those allowed within the conventional set-up,
i.e. the pure standard FHI without the
complementary MI (compare with
Refs.~\refcite{jp,sstad}).

\item
As $\kappa$ increases above 0.01, the SUGRA
corrections become more and more significant.

\item
As $\kappa$ decreases below about 0.015
[0.042], the constraint from the lower
[upper] bound on $n_{\rm s}$ ceases to
restrict the parameters, since it is
overshadowed by the lower [upper] bound on
$N_{\rm HI*}$ [$m_s/H_s$] in Eq.~(\ref{ten})
[Eq.~(\ref{nin})].

\item
The running $dn_{\rm s}/d\ln k$ of the spectral
index satisfies comfortably the bound
in Eq.~(\ref{dnsk}) in the largest part of the
regions allowed by the other constraints, whereas
$-0.005\gtrsim dn_{\rm s}/d\ln k\gtrsim -0.01$ in
a very limited part of these regions.

\item
For $n_{\rm s}=0.958$, we obtain $0.004\lesssim
\kappa\lesssim0.14$, $0.79\lesssim v_{_G}/
(10^{16}~{\rm GeV})\lesssim 1.08$, and
$-0.002\gtrsim dn_{\rm s}/d\ln k\gtrsim-0.01$.
Also, $10\lesssim N_{\rm HI*}\lesssim 21.7$,
$35\gtrsim N_{\rm MI} \gtrsim 24$, and
$0.64\lesssim m_s/H_s\lesssim 0.77$.

\end{romanlist}

\subsection{Shifted FHI}

In Fig.~\ref{shth}, we delineate the
regions allowed by Eqs.~(\ref{nswmap}),
(\ref{Prob})--(\ref{nin}), and (\ref{random}) in
the {(a)} $\kappa-v_{_G}$, {(b)}
$\kappa-m_s/H_s$, {(c)} $\kappa-N_{\rm HI*}$, and
{(d)} $\kappa-N_{\rm MI}$ plane for shifted FHI
with $M_{\rm S}=5\times10^{17}~{\rm GeV}$. Some
observations are in order:

\begin{romanlist}[(ii)]

\item
The lower [upper] bound on $N_{\rm HI*}$
[$m_s/H_s$] in Eq.~(\ref{ten})
[Eq.~(\ref{nin})] gives a lower [upper] bound
on $v_{_G}$ for each $\kappa$, in contrast to
the case of standard FHI.

\item
The results on $m_s/H_s$, $N_{\rm HI*}$, and
$N_{\rm MI}$ are quite similar to those
obtained in the case of standard FHI.

\item
The common magnitude $v_{_G}$ of the VEVs of
$\bar\Phi$ and $\Phi$ comes out considerably
larger than in the case of standard FHI and can
be put equal to the SUSY GUT scale. Some key
inputs and outputs for the interesting case with
$v_{_G}=M_{\rm GUT}$ and $n_{\rm s}=0.958$ are
presented in Table~\ref{tabsh}.

\end{romanlist}

\begin{figure}[t!]
\centerline{
\psfig{file=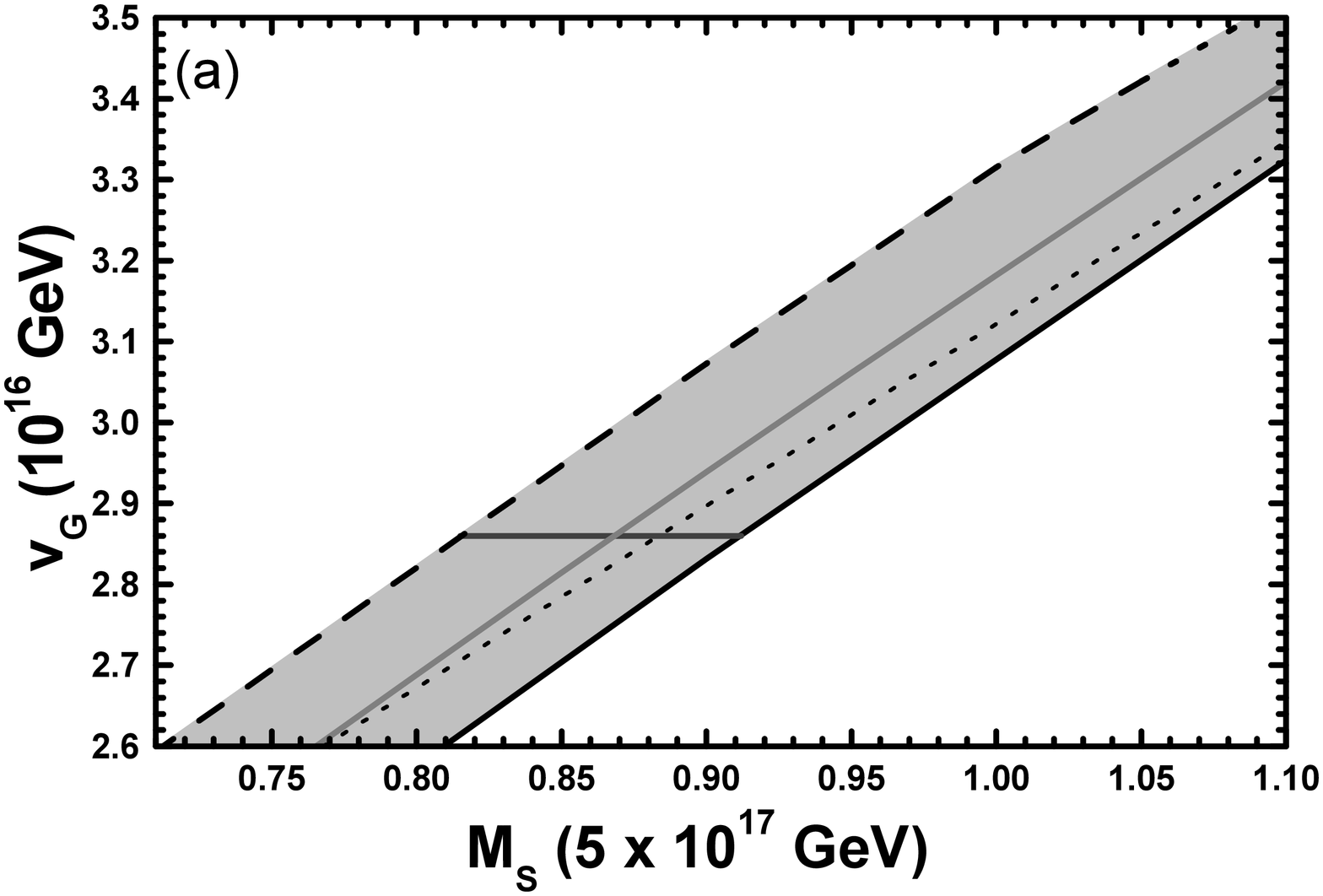,width=4.4cm,angle=-90}
\psfig{file=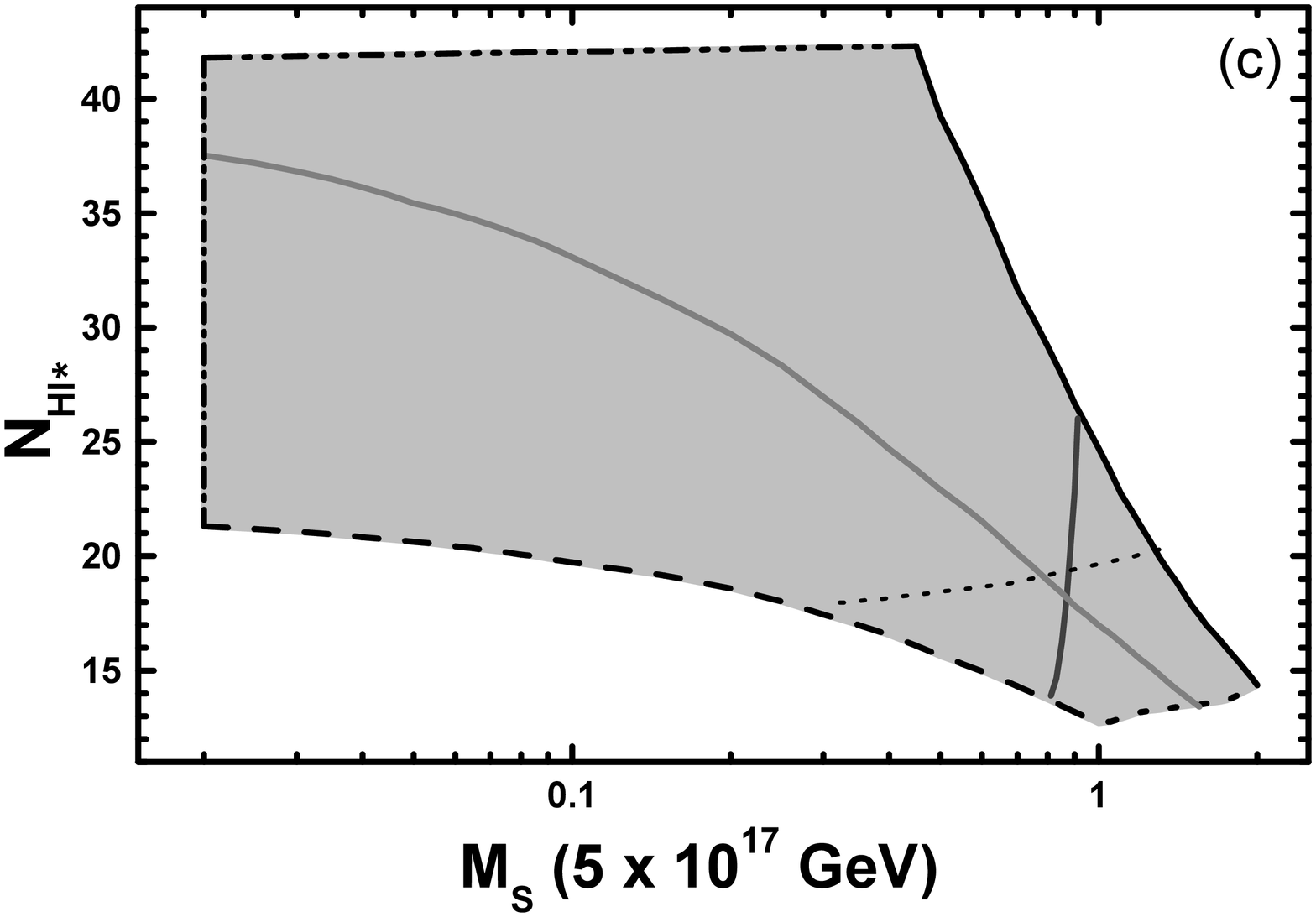,width=4.4cm,angle=-90}}
\centerline{
\psfig{file=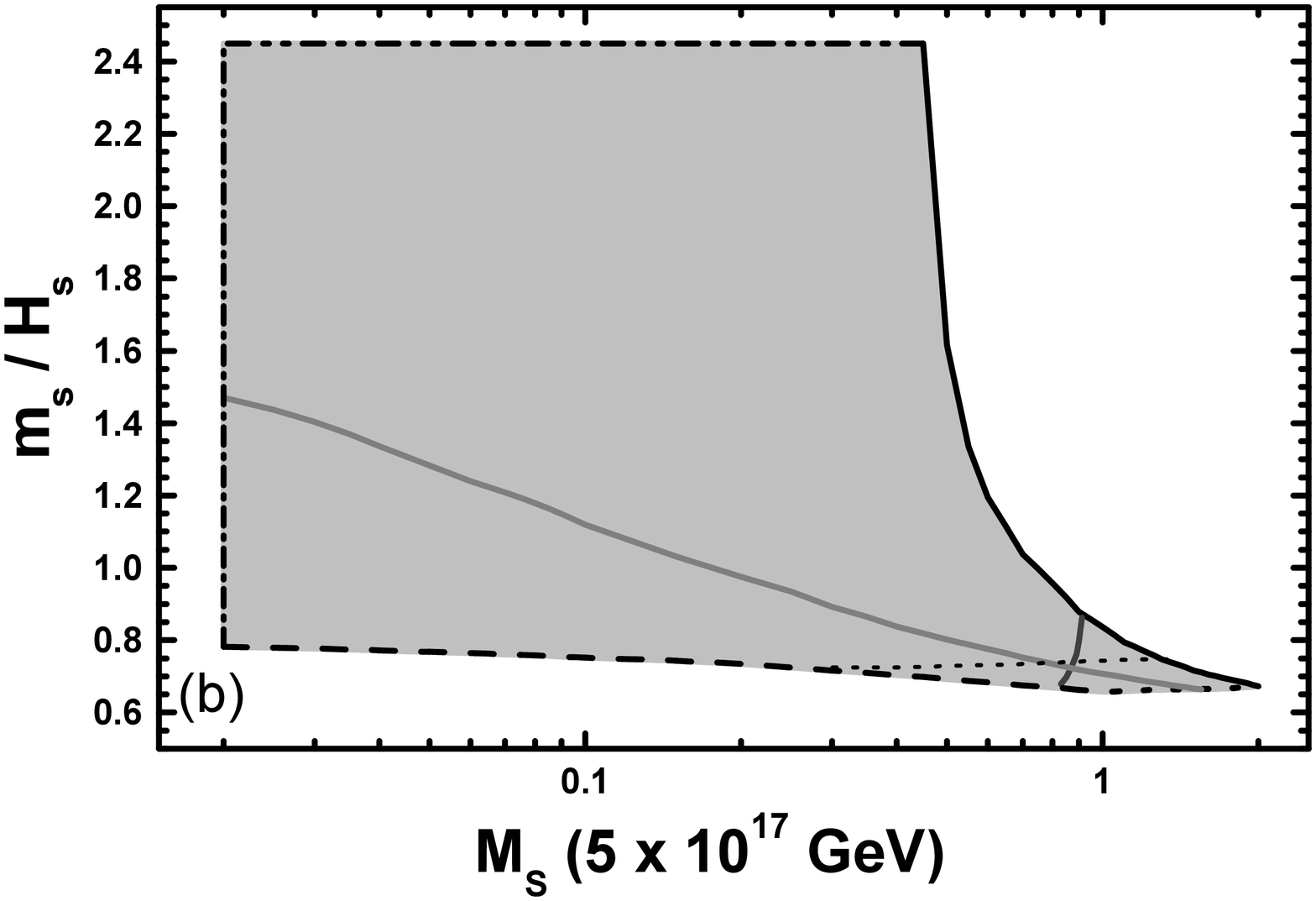,width=4.4cm,angle=-90}
\psfig{file=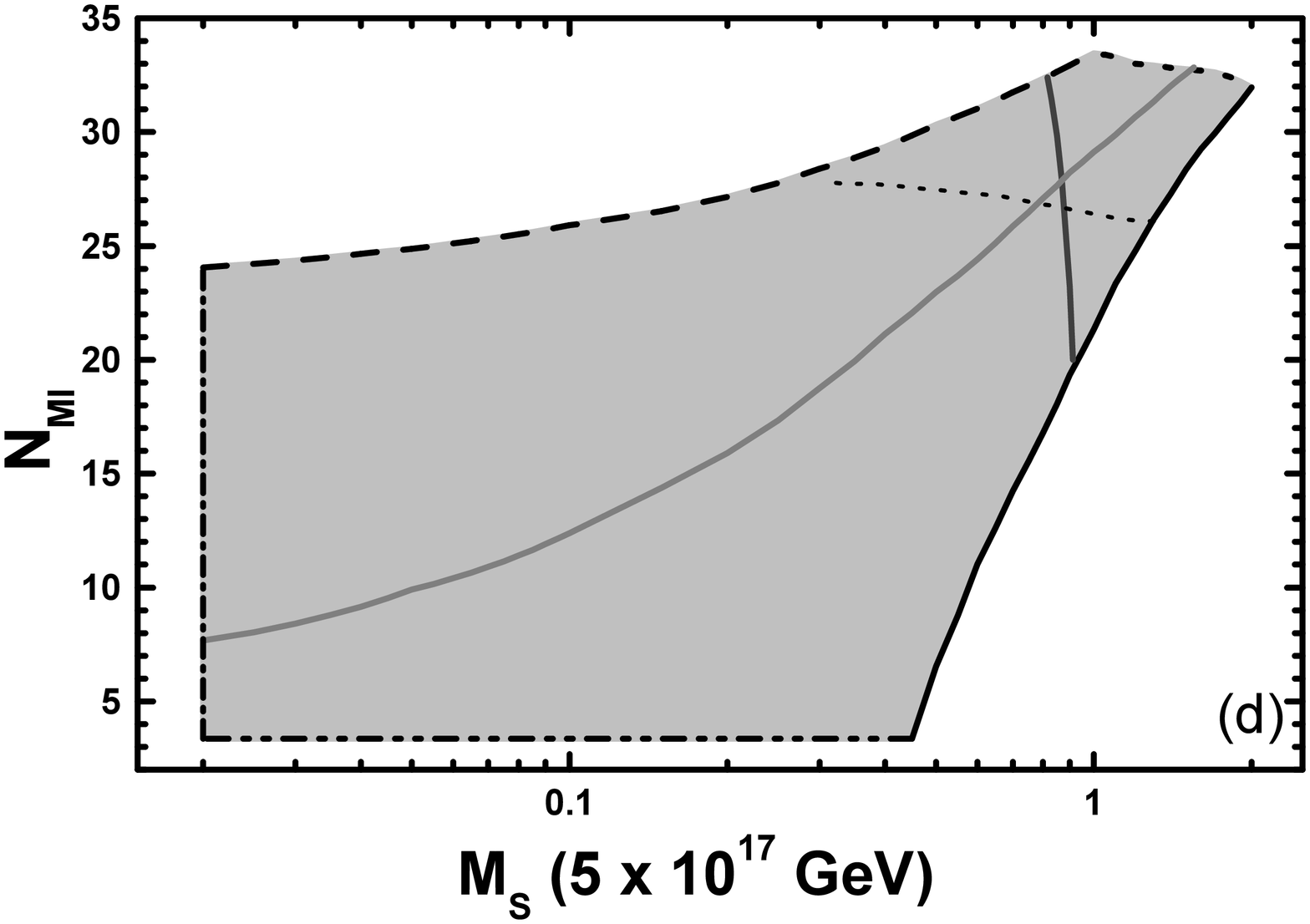,width=4.4cm,angle=-90}}
\vspace*{8pt}
\caption{Allowed regions in the {(a)}
$M_{\rm S}-v_{_{G}}$, {(b)} $M_{\rm S}-m_s/H_s$,
{(c)} $M_{\rm S}-N_{\rm HI*}$, and {(d)}
$M_{\rm S}-N_{\rm MI}$ plane for smooth FHI.
Same notation as in Fig.~\ref{shth}. We included
small $M_{\rm S}$'s of less physical interest
just to show the effect of the constraints.}
\label{smth}
\end{figure}

\subsection{Smooth FHI}

In Fig.~\ref{smth}, we present the regions
allowed by Eqs.~(\ref{nswmap}),
(\ref{Prob})--(\ref{nin}), and (\ref{random}) in
the {(a)} $M_{\rm S}-v_{_G}$, {(b)}
$M_{\rm S}-m_s/H_s$,
{(c)} $M_{\rm S}-N_{\rm HI*}$, and {(d)}
$M_{\rm S}-N_{\rm MI}$ plane for smooth FHI.
We observe the following:

\begin{romanlist}[(ii)]

\item
The SUGRA corrections play an important role
for every $M_{\rm S}$ in the allowed regions
of Fig.~\ref{smth}.

\item
In contrast to standard and shifted FHI,
$|dn_{\rm s}/d\ln k|$ is considerably enhanced
with $-0.005\gtrsim dn_{\rm s}/d\ln k\gtrsim
-0.01$ holding in a sizable portion of the
parameter space for $v_{_G}\sim M_{\rm GUT}$.

\item
Unlike the cases of standard and shifted FHI,
the constraint of Eq.~(\ref{ten}) does not
restrict the parameters.

\item
Similarly to the case of shifted FHI, we can
find an acceptable solution fixing $n_{\rm s}
=0.958$ and $v_{_G}=M_{\rm GUT}$. Some key
inputs and outputs of this solution are
arranged in Table~\ref{tabsh}.

\end{romanlist}

\begin{table}[t]
\tbl{Input and output parameters for our scenario
with shifted ($M_{\rm S}=5\times 10^{17}~
{\rm GeV}$) or smooth FHI for $n_{\rm s}=0.958$
and $v_{_G}=M_{\rm GUT}$.}
{\begin{tabular}{lr@{\hspace{0.5cm}}clr}
\toprule
\multicolumn{2}{c}{Shifted FHI}&&
\multicolumn{2}{c}{Smooth FHI}\\
\colrule
$\sigma_*~(10^{16}~{\rm GeV})$ & $2.2$&&
$\sigma_*~(10^{16}~{\rm GeV})$ & $23.53$\\
$\kappa$ & $0.01$&&$M_{\rm S}~(5\times 10^{17}~
{\rm GeV})$ &$0.87$\\ \colrule
$M~(10^{16}~{\rm GeV})$& $2.35$&&$\mu_{\rm S}
~(10^{16}~{\rm GeV})$& $0.188$\\
$1/\xi$ &  4.54&&$\sigma_{\rm f}~(10^{16}~
{\rm GeV})$&$13.42$ \\
$N_{\rm HI*}$ &  $21$&&$N_{\rm HI*}$ & $18$\\
$dn_{\rm s}/d\ln k$ &  $-0.0018$&&
$dn_{\rm s}/d\ln k$ &$-0.0055$\\ \colrule
$N_{\rm MI}$ & $24.3$&&$N_{\rm MI}$ & $27.8$\\
$m_s/H_s$ & $0.77$&&$m_s/H_s$ & $0.72$\\
\botrule
\end{tabular}
\label{tabsh}}
\end{table}

\section{Conclusions}
\label{con}

We presented a recently proposed\cite{mhin}
cosmological scenario tied to two bouts of
inflation: a GUT scale FHI which reproduces the
current data on $P_{\cal R}$ and $n_{\rm s}$
within the power-law $\Lambda$CDM cosmological
model and generates a restricted number of
e-foldings $N_{\rm HI*}$ followed by an
intermediate scale MI which generates the
residual number of e-foldings. We assumed that
the inflaton of MI is a string axion which
remains naturally almost massless during FHI
(and the inter-inflationary period). We
considered extra restrictions on the
parameters of the model originating from the
following:

\begin{romanlist}[(ii)]

\item
The resolution of the horizon and flatness
problems of the standard hot big bang cosmology.

\item
The requirements that FHI lasts long enough to
generate the observed primordial fluctuations on
all the cosmological scales and that these scales
are not re-processed by the subsequent MI.

\item
The limit on the running of the spectral index.

\item
The naturalness of MI.

\item
The homogeneity of the present universe.

\item
The complete randomization of the string axion
during FHI.

\end{romanlist}

Fixing the spectral index to its central value,
we concluded the following:

\begin{romanlist}[(ii)]

\item
In the case of standard FHI, relatively large
values of the dimensionless parameter $\kappa$
and the GUT breaking VEV $v_{_G}$ are required
and $10\lesssim N_{\rm HI*}\lesssim 21.7$.

\item
In the shifted [smooth] FHI case, identification
of the GUT breaking VEV with the SUSY GUT scale
is possible provided that $N_{\rm HI*}\simeq 21$
[$N_{\rm HI*}\simeq 18$].

\end{romanlist}

\par
In all three versions of hybrid inflation studied
here with $n_{\rm s}$ near its central value and,
in the smooth FHI case, $v_{_G}\sim M_{\rm GUT}$,
MI of the slow-roll type with
$m_s/H_s\sim 0.6-0.8$ and a very mild tuning (of
order 0.01) of the initial value of the string
axion produces the additional number of
e-foldings required for solving the horizon and
flatness problems of standard hot big bang
cosmology. Therefore, MI complements successfully
FHI.

\par
Note that MI naturally assures a low reheat
temperature. As a consequence, baryogenesis is
made more difficult. In particular,
thermal\cite{thermallepto} or
non-thermal\cite{nonthermallepto} leptogenesis
won't work since the reheat temperature is very
low for the non-perturbative electroweak
sphalerons to operate. However, it is not
impossible to achieve\cite{benakli} adequate
baryogenesis within a larger scheme
with (large) extra dimensions. Let us also
mention that, due to the presence of MI, the
gravitino constraint\cite{gravitino} on the
reheat temperature of FHI and the potential
topological defect problem of standard FHI can
be significantly relaxed or completely evaded.
Our set-up is beneficial for MI too, since, due
to its low inflationary scale, this model cannot
account for the observed primordial fluctuations
(unless a special mechanism\cite{pngbcd} is
employed).

\section*{Acknowledgments}

We thank K. Dimopoulos and R. Trotta for
discussions. This work has been supported by the
European Union under the contracts
MRTN-CT-2004-503369 and HPRN-CT-2006-035863 as
well as by the PPARC research grant PP/C504286/1.

\def\aipcp#1#2#3{{\it AIP Conf. Proc.}
{\bf #1},~#3~(#2)}
\def\ijmp#1#2#3{{\it Int. Jour. Mod. Phys.}
{\bf #1},~#3~(#2)}
\def\plb#1#2#3{{\it Phys. Lett. B }{\bf #1},~#3~(#2)}
\def\zpc#1#2#3{{\it Z. Phys. C }{\bf #1},~#3~(#2)}
\def\prl#1#2#3{{\it Phys. Rev. Lett.}
{\bf #1},~#3~(#2)}
\def\rmp#1#2#3{{\it Rev. Mod. Phys.}
{\bf #1},~#3~(#2)}
\def\prep#1#2#3{{\it Phys. Rep. }{\bf #1},~#3~(#2)}
\def\prd#1#2#3{{\it Phys. Rev. D }{\bf #1},~#3~(#2)}
\def\npb#1#2#3{{\it Nucl. Phys. }{\bf B#1},~#3~(#2)}
\def\npbps#1#2#3{{\it Nucl. Phys. B, Proc. Suppl.}
{\bf #1},~#3~(#2)}
\def\mpl#1#2#3{{\it Mod. Phys. Lett.}
{\bf #1},~#3~(#2)}
\def\arnps#1#2#3{{\it Annu. Rev. Nucl. Part. Sci.}
{\bf #1},~#3~(#2)}
\def\sjnp#1#2#3{{\it Sov. J. Nucl. Phys.}
{\bf #1},~#3~(#2)}
\def\app#1#2#3{{\it Acta Phys. Polon.}
{\bf #1},~#3~(#2)}
\def\rnc#1#2#3{{\it Riv. Nuovo Cim.}
{\bf #1},~#3~(#2)}
\def\ap#1#2#3{{\it Ann. Phys. }{\bf #1},~#3~(#2)}
\def\ptp#1#2#3{{\it Prog. Theor. Phys.}
{\bf #1},~#3~(#2)}
\def\apjl#1#2#3{{\it Astrophys. J. Lett.}
{\bf #1},~#3~(#2)}
\def\n#1#2#3{{\it Nature }{\bf #1},~#3~(#2)}
\def\apj#1#2#3{{\it Astrophys. J.}
{\bf #1},~#3~(#2)}
\def\anj#1#2#3{{\it Astron. J. }{\bf #1},~#3~(#2)}
\def\mnras#1#2#3{{\it MNRAS }{\bf #1},~#3~(#2)}
\def\grg#1#2#3{{\it Gen. Rel. Grav.}
{\bf #1},~#3~(#2)}
\def\s#1#2#3{{\it Science }{\bf #1},~#3~(#2)}
\def\baas#1#2#3{{\it Bull. Am. Astron. Soc.}
{\bf #1},~#3~(#2)}
\def\ibid#1#2#3{{\it ibid. }{\bf #1},~#3~(#2)}
\def\cpc#1#2#3{{\it Comput. Phys. Commun.}
{\bf #1},~#3~(#2)}
\def\astp#1#2#3{{\it Astropart. Phys.}
{\bf #1},~#3~(#2)}
\def\epjc#1#2#3{{\it Eur. Phys. J. C}
{\bf #1},~#3~(#2)}
\def\nima#1#2#3{{\it Nucl. Instrum. Meth. A}
{\bf #1},~#3~(#2)}
\def\jhep#1#2#3{{\it J. High Energy Phys.}
{\bf #1},~#3~(#2)}
\def\jcap#1#2#3{{\it J. Cosmol. Astropart. Phys.}
{\bf #1},~#3~(#2)}
\def\jpa#1#2#3{{\it J. Phys. A}
{\bf #1},~#3~(#2)}
\def\lnp#1#2#3{{\it Lect. Notes Phys.}
{\bf #1},~#3~(#2)}
\def\jpcs#1#2#3{{\it J. Phys. Conf. Ser.}
{\bf #1},~#3~(#2)}
\def\jetpl#1#2#3{{\it JETP Lett.}
{\bf #1},~#3~(#2)}
\def\jetpsp#1#2#3{{\it JETP (Sov. Phys.)}
{\bf #1},~#3~(#2)}
\def\astroph#1{{astro-ph/}{#1}}
\def\hepph#1{{hep-ph/}{#1}}
\def\hepth#1{{hep-th/}{#1}}

\end{document}